\documentclass[twocolumn,showpacs,amsmath,amssymb,pre,a4paper,floatfix,showkeys]{revtex4}
\usepackage{dcolumn}
\usepackage{graphicx}
\usepackage{epsfig}

\begin{document}

\title{Event-by-event simulation of quantum phenomena\footnote{V Brazilian Meeting on Simulational Physics, Ouro Preto, 2007.
Brazilian Journal of Physics (in press).}}

\author{H. De Raedt}
\email{h.a.de.raedt@rug.nl}
\affiliation{Department of Applied Physics, Zernike Institute for Advanced Materials,
University of Groningen, Nijenborgh 4, NL-9747 AG Groningen, The Netherlands}
\date{\today}


\begin{abstract}
In this talk, I discuss recent progress in the development of simulation algorithms that
do not rely on any concept of quantum theory but are nevertheless capable of reproducing
the averages computed from quantum theory through an event-by-event simulation.
The simulation approach is illustrated by applications to single-photon
Mach-Zehnder interferometer experiments and Einstein-Podolsky-Rosen-Bohm experiments with photons.
\end{abstract}

\maketitle

\def\sumprime{\mathop{{\sum}'}}
\def\DLM{DLM}
\def\DLMS{DLMs}
\def\Eq#1{(\ref{#1})}

\section{Introduction}
\label{Introduction}

Computer simulation is widely regarded as complementary to theory and experiment~\cite{LAND00}.
The standard procedure is to start from one or more basic equations of physics and
to apply existing or invent new algorithms to solve these equations.
This approach has been highly successful for a wide variety of problems
in science and engineering.
However, there are a number of physics problems, very fundamental ones,
for which this approach fails, simply because there are no basic
equations to start from.

Indeed, as is well-known from the early days in the development of quantum theory,
quantum theory has nothing to say about individual events~\cite{BOHM51,HOME97,BALL03}.
Reconciling the mathematical formalism that does not describe individual events
with the experimental fact that each observation yields a definite outcome
is referred to as the quantum measurement paradox and is
the most fundamental problem in the foundation of quantum theory~\cite{HOME97}.

If computer simulation is indeed a third methodology,
it should be possible to simulate quantum phenomena on an event-by-event basis.
For instance, it should be possible to simulate that we can see, with our own eyes, how
in a two-slit experiment with single electrons, an interference pattern
appears after a considerable
number of individual events have been recorded by the detector~\cite{TONO98}.

In view of the quantum measurement paradox, it is unlikely
that we can find such a simulation algorithm by limiting our thinking to
the framework of quantum theory.
Of course, we could simply use pseudo-random numbers to generate events according to the probability distribution
that is obtained by solving the time-independent Schr{\"o}dinger equation.
However, that is not what we mean when we say that within the framework of quantum theory,
there is little hope to find an algorithm that simulates the individual events
and reproduces the expectation values obtained from quantum theory.
The challenge is to find algorithms that simulate, event-by-event,
the experimental observations that, for instance, interference patterns appear only after a considerable
number of individual events have been recorded by the detector~\cite{GRAN86,TONO98}, without
first solving the Schr{\"o}dinger equation.

In a number of recent papers~\cite{RAED05b,RAED05c,RAED05d,MICH05,RAED06z,MICH06z,RAED06c,RAED07a,RAED07c},
we have demonstrated that locally-connected networks of processing units
can simulate event-by-event, the single-photon beam splitter and Mach-Zehnder interferometer
experiments of Grangier \textit{et al}.~\cite{GRAN86}.
Furthermore, we have shown that this approach can be generalized to simulate
universal quantum computation by an event-by-event process~\cite{RAED05c,MICH05,MICH06z},
and that it can be used to simulate real Einstein-Podolsky-Rosen-Bohm (EPRB) experiments~\cite{RAED06c,RAED07a,RAED07c}.
Therefore, at least in principle, our approach can be used to simulate
all wave interference phenomena and many-body quantum systems using particle-like processes only.
Our work suggests that we may have discovered a procedure to simulate quantum phenomena
using event-based processes that satisfy Einstein's criterion of local causality.

This talk is not about interpretations or extensions of quantum theory.
The fact that there exist simulation algorithms that reproduce the results of quantum theory
has no direct implications on the foundations of quantum theory:  These algorithms describe the process of generating events
on a level of detail about which quantum theory has nothing to say~\cite{HOME97,BALL03}.
The average properties of the data may be in perfect agreement with quantum theory but
the algorithms that generate such data are outside of the scope of what quantum theory can describe.
This may sound a little strange but it may not be that strange if one recognizes that probability theory
does not contain nor provides an algorithm to generate the values of the random variables either,
which in a sense, is at the heart of the quantum measurement paradox~\cite{RAED07c}.

\begin{figure*}[t]
\begin{center}
\includegraphics[width=16cm]{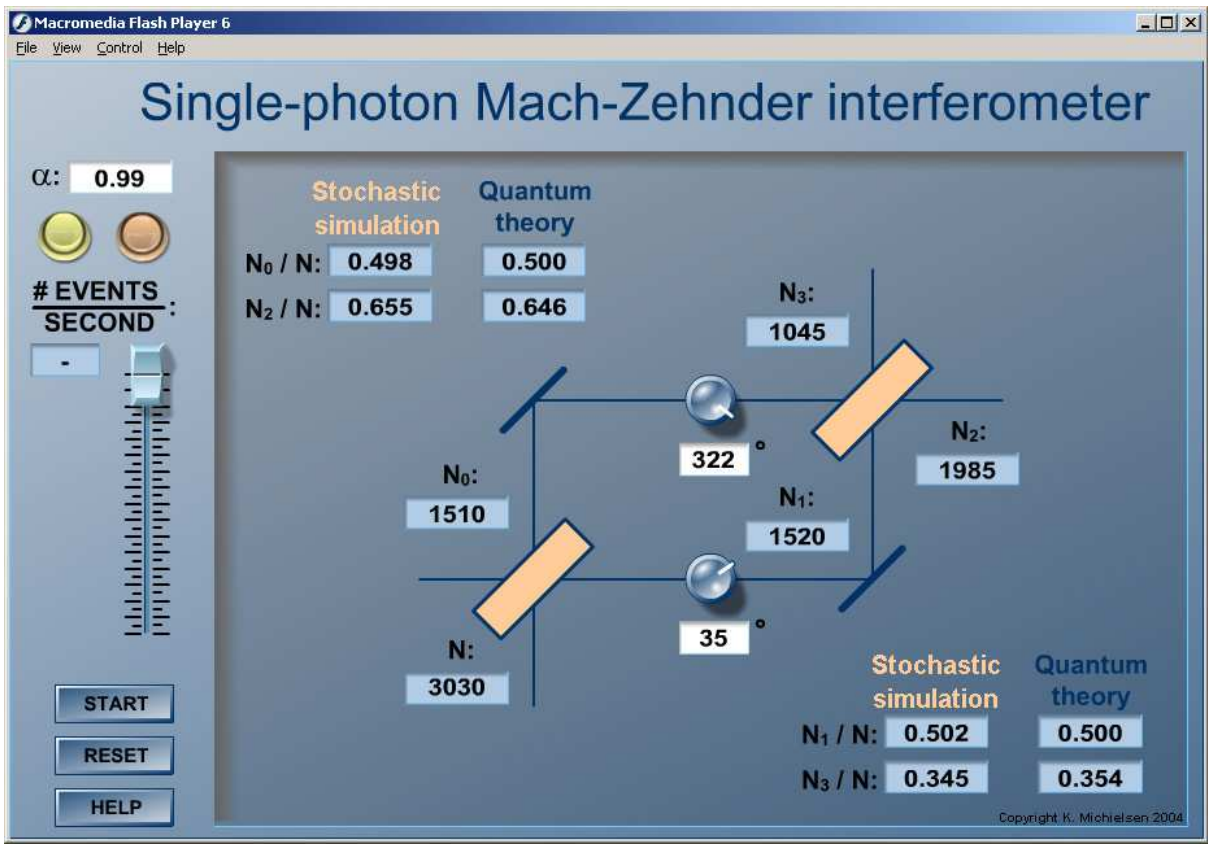}
\caption{(color online)
Snapshot of an interactive event-by-event simulator
of a Mach-Zehnder interferometer~\cite{MZIdemo}.
The main panel shows the layout of the interferometer.
Particles emerge from a source (not shown)
located at the bottom of the left-most vertical line.
After leaving the first beam splitter in either
the vertical or horizontal direction,
the particles experience time delays
that are specified by the controls on the lines.
In this example, the time delays correspond
to the phase shifts $\phi_0=35^\circ$ and
$\phi_1=322^\circ$ in the wave mechanical description.
The thin, $45^\circ$-tilted lines act as perfect mirrors.
When a particle leaves the system at the top right,
it adds to the count of either detector $N_2$ or $N_3$.
Additional detectors ($N_0$, $N_1$)
count the number of particles on the corresponding lines.
The other cells give the ratio of the detector counts
to the total number of particles (messages) processed
and also the corresponding probability of
the quantum mechanical description.
At any time, the user can choose between a
strictly deterministic and a stochastic event-by-event
simulation by pressing the buttons at the top of the control panel.
}
\label{onemzi}
\end{center}
\end{figure*}

\section{Single-photon Mach-Zehnder interferometer}{\label{sec5}}

Figure~\ref{onemzi} shows the schematic diagram of a Mach-Zehnder interferometer~\cite{BORN64}.
From Maxwell's theory of classical electrodynamics it follows that the intensity
of light recorded by detectors $N_2$ and $N_3$ is proportional to $\cos^2\phi/2$ and
$\sin^2\phi/2$, respectively~\cite{BORN64}.
Here $\phi=\phi_1-\phi_2$ is the phase difference that
expresses the fact that depending on which path the light takes to travel from
the first to the second beam splitter, the optical path length may be different~\cite{BORN64}.

It is an experimental fact that when the Mach-Zehnder interferometer experiment
is carried out with one photon at a time,
the number of individual photons recorded
by detectors $N_2$ and $N_3$ is proportional to $\cos^2\phi/2$ and
$\sin^2\phi/2$~\cite{GRAN86}, in agreement with classical electrodynamics.
In quantum physics~\cite{QuantumTheory}, single-photon experiments with one beam splitter
provide direct evidence for the particle-like behavior of photons.
The wave mechanical character appears when one performs
interference experiments with individual particles~\cite{GRAN86,HOME97}.
Quantum physics ``solves'' this logical contradiction by introducing
the concept of particle-wave duality~\cite{HOME97}.

In this section, we describe a system that does not build on any concept of quantum theory yet
displays the same interference patterns as those observed in single-photon
Mach-Zehnder interferometer experiments~\cite{GRAN86}.
The basic idea is to describe (quantum) processes in terms of events, messages,
and units that process these events and messages.
In the experiments of Grangier et al.~\cite{GRAN86},
the photon carries the message (a phase),
an event is the arrival of a photon at one of the input ports
of a beam splitter, and the beam splitters are the processing units.
In experiments with single photons, there is no way other than through
magic, by which a photon can communicate directly with another photon.
Thus, it is not difficult to imagine that if we want a system to
exhibit some kind of interference, the communication among
successive photons should take place in the beam splitters.

In this talk, we consider the simplest processing unit that
is adequate for our purpose, namely a standard linear adaptive filter~\cite{RAED05d}.
The processing unit receives a message through one of its input ports,
processes the message according to some rule (see later), and
sends a message (carried by the messenger, that is a photon)
through an output port that it selects using a pseudo-random number,
drawn from a distribution that is determined
by the current state of the processing unit.
Other, more complicated processing units that operate in a fully deterministic
manner are described elsewhere~\cite{RAED05b,MICH05}.
Although the sequence of events that the different types of processing
units produce can be very different, the quantities that
are described by quantum theory, namely the averages,
are the same.
The essential feature of all these processing units is their ability
to learn from the events they process.
Processing units that operate according to this principle
will be referred to as deterministic learning machines (DLMs)~\cite{RAED05b,MICH05}.

By connecting an output channel to the input channel of another \DLM, we can build
networks of \DLMS.
As the input of a network receives an event,
the corresponding message is routed through the network while it is being
processed and eventually a message appears at one of the outputs.
At any given time during the processing,
there is only one output-to-input connection in the network that is actually carrying a message.
The \DLMS\ process the messages in a sequential manner and communicate with each other
by message passing.
There is no other form of communication between different \DLMS.
The parts of the processing units and network map one-to-one
on the physical parts of the experimental setup
and only simple geometry is used to construct the simulation algorithm.
Although networks of \DLMS\ can be viewed as networks that are capable of unsupervised learning,
they have little in common with neural networks.
It obvious that this simulation approach
satisfies Einstein's criteria of realism and local causality~\cite{HOME97}.

\subsection{Beam splitter}
\label{BS}

\begin{figure}[t]
\begin{center}
\includegraphics[width=8.5cm]{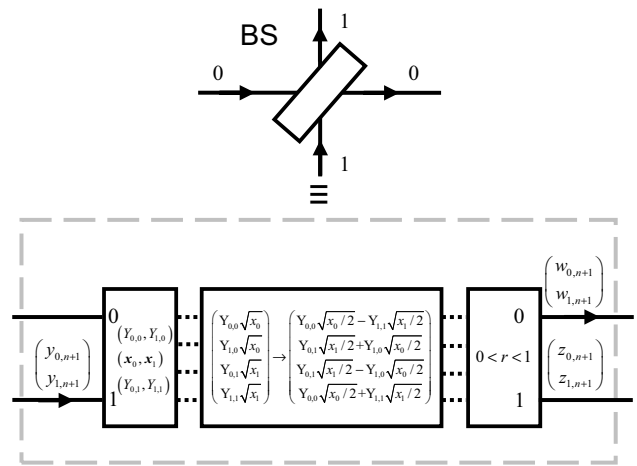}
\caption{
Diagram of a \DLM\ that performs an event-by-event
simulation of a single-photon beam splitter (BS)~\cite{MZIdemo}.
The solid lines represent the input and output channels of the BS.
Dashed lines indicate the flow of data within the BS.
}
\label{figbs}
\label{one-bs}
\end{center}
\end{figure}

Figure~\ref{figbs} shows the schematic diagram of a \DLM\ that simulates
a beam splitter, event-by-event.
We label events by a subscript $n\ge0$.
At the $(n+1)$th event, the \DLM\ receives a message on either input channel
0 or 1, never on both channel simultaneously.
Every message consists of a two-dimensional unit vector ${\bf y}_{n+1}=(y_{1,n+1},y_{2,n+1})$.
This vector represents the phase of the event that occurs on channel 0 (1).
Although it would be sufficient to use the phase itself as the message,
in practice it is more convenient to work with the cosine
($y_{1,n+1}$) and sine ($y_{2,n+1}$) of the phase.

The first stage of the \DLM\ (see Fig.~\ref{figbs})
stores the message ${\bf y}_{n+1}$ in its internal register ${\bf Y}_k$.
Here, $k=0$ (1) if the event occurred on channel 0 (1).
The first stage also has an internal two-dimensional vector
${\bf x}=(x_{0},x_{1})$
with the additional constraints that $x_{i}\ge0$ for $i=0,1$
and that $x_{0} +x_{1}=1$.
As we only have two input channels, the latter constraint
can be used to recover $x_1$ from the value of $x_0$.
We prefer to work with internal vectors that have as many elements
as there are input channels.
After receiving the $(n+1)$-th event on input channel $k=0,1$
the internal vector is updated according to the rule
\begin{eqnarray}
x_{i,n+1}&=&\alpha x_{i,n} + 1-\alpha
\quad\hbox{if}\quad i=k,
\nonumber \\
x_{i,n+1}&=&\alpha x_{i,n}
\quad\hbox{if}\quad i\not=k,
\label{HYP1}
\end{eqnarray}
where $0<\alpha<1$ is a parameter.
By construction $x_{i,n+1}\ge0$ for $i=0,1$ and $x_{0,n+1} +x_{1,n+1}=1$.
Hence the update rule Eqs.~\Eq{HYP1} preserves the constraints
on the internal vector.
Obviously, these constraints are necessary if we want to interpret
the $x_{k,n}$ as (an estimate of) the probability
for the occurrence of an event of type $k$.

The second stage of the \DLM\
takes as input the values stored in the registers ${\bf Y}_0$, ${\bf Y}_1$, {\bf x}
and transforms this data according to the rule
\begin{eqnarray}
\frac{1}{\sqrt{2}}
\left(
\begin{array}{c}
Y_{0,0}\sqrt{x_0}-Y_{1,1}\sqrt{x_1}\\
Y_{0,1}\sqrt{x_1}+Y_{1,0}\sqrt{x_0}\\
Y_{0,1}\sqrt{x_1}-Y_{1,0}\sqrt{x_0}\\
Y_{0,0}\sqrt{x_0}+Y_{1,1}\sqrt{x_1}
\end{array}
\right)
{\longleftarrow}
\left(
\begin{array}{c}
Y_{0,0}\sqrt{x_0}\\
Y_{1,0}\sqrt{x_0}\\
Y_{0,1}\sqrt{x_1}\\
Y_{1,1}\sqrt{x_1}
\end{array}
\right),
\label{BS1}
\end{eqnarray}
where we have omitted the event label $(n+1)$ to simplify the notation.
Note that the second subscript of the ${\bf Y}$-register refers to the
type of input event.

The third stage of the \DLM\ in Fig.~\ref{figbs}
responds to the input event by sending a message
${\bf w}_{n+1}=( Y_{0,0}\sqrt{x_0}-Y_{1,1}\sqrt{x_1},Y_{0,1}\sqrt{x_1}+Y_{1,0}\sqrt{x_0})/\sqrt{2}$
through output channel 0 if
$w_{0,n+1}^2+w_{1,n+1}^2>r$
where $0<r<1$ is a uniform random number.
Otherwise the back-end sends the message
${\bf z}_{n+1}=( Y_{0,1}\sqrt{x_1}-Y_{1,0}\sqrt{x_0},Y_{0,0}\sqrt{x_0}+Y_{1,1}\sqrt{x_1})/\sqrt{2}$
through output channel 1.
Finally, for reasons of internal consistency of the simulation method, it is necessary
to replace
${\bf w}_{n+1}$  by ${\bf w}_{n+1}/\Vert{\bf w}_{n+1}\Vert$ or
${\bf z}_{n+1}$  by ${\bf z}_{n+1}/\Vert{\bf z}_{n+1}\Vert$
such that the output message is represented by a unit vector.

It is almost trivial to perform a computer simulation of the \DLM\
model of the beam splitter and convince oneself that it reproduces
all the results of quantum theory for this device~\cite{RAED05d}.
With only a little more effort, it can be shown that the input-output behavior of the
\DLM\ is, on average, the same as that of the (ideal) beam splitter.

According to quantum theory, the probability amplitudes ($b_0,b_1)$
of the photons in the output modes 0 and 1 of a
beam splitter (see Fig.~\ref{figbs}) are given by~\cite{BAYM74,GRAN86,RARI97}
\begin{eqnarray}
\left(
\begin{array}{c}
b_0\\
b_1
\end{array}
\right)
=
\frac{1}{\sqrt{2}}
\left(
\begin{array}{c}
a_0+ia_1\\
a_1+ia_0
\end{array}
\right)
=
\frac{1}{\sqrt{2}}
\left(
\begin{array}{cc}
1&i\\
i&1
\end{array}
\right)
\left(
\begin{array}{c}
a_0\\
a_1
\end{array}
\right),
\label{BS3}
\end{eqnarray}
where the presence of photons in the input modes 0 or 1 is represented
by the probability amplitudes ($a_0,a_1)$~\cite{BAYM74,GRAN86,RARI97}.
From Eq.~\ref{BS3}, it follows that the intensities recorded
by detectors $N_0$ and $N_1$ is given by
\begin{eqnarray}
|b_0|^2=\frac{1+2\sqrt{p_0(1-p_0)}\sin(\psi_0-\psi_1)}{2},
\nonumber \\
|b_1|^2=\frac{1-2\sqrt{p_0(1-p_0)}\sin(\psi_0-\psi_1)}{2}.
\label{BS4}
\end{eqnarray}

On the other hand, the formal solution of Eq.~\Eq{HYP1} reads
\begin{equation}
{\bf x}_{n}=\alpha^n {\bf x}_0 + (1-\alpha)\sum_{i=0}^{n-1}\alpha^{n-1-i}{\bf v}_{i+1}
,
\label{HYP2}
\end{equation}
where ${\bf x}_n=(x_{0,n},x_{1,n})$,
and ${\bf x}_0$ denotes the initial value of the internal vector.
The input events are represented by the vectors
${\bf v}_{n+1}=(1,0)^T$ or ${\bf v}_{n+1}=(0,1)^T$ if the ${n+1}$-th
event occurred on channel 0 or 1, respectively.
Let $p_0$ ($(1-p_0)$) be the probability for an input event of type 0 (1).
Taking the average of Eq.\Eq{HYP2} over many events and using $0<\alpha<1$,
we find that for large $n$, ${\bf x}_n\approx(p_0,1-p_0)^T$.
Therefore the first stage of the DLM ``learns'' the probabilities
for events 0 and 1 by processing these events in a sequential manner.
The parameter $0<\alpha<1$ controls the learning process.

Using two complex numbers instead of four real numbers
that enter Eq.~\Eq{BS1}, identification of
$a_0$ with $Y_{0,0}\sqrt{x_0}+i Y_{1,0}\sqrt{x_0}$
and $a_1$ with $Y_{0,1}\sqrt{x_1}+i Y_{1,1}\sqrt{x_1}$
shows that the transformation stage plays the
role of the matrix-vector multiplication
in Eq.\Eq{BS3}. By construction, the output stage
receives as input the four real numbers that
correspond to $b_0$ and $b_1$.
Thus, after the \DLM\ has reached the stationary state,
it will distribute events over its output channels according to
Eq.\Eq{BS4}. Of course, this reasoning is firmly supported
by extensive simulations~\cite{RAED05d,RAED05b}.

One may wonder what learning machines have to do with the (wave) mechanical models that we are accustomed to
in physics. First, one should keep in mind that the approach that I describe in this talk is capable of giving
a rational, logically consistent description of event-based phenomena that cannot be incorporated in
a wave mechanical theory without adding logically incompatible concepts such as the wave function collapse~\cite{HOME97}.
Second, the fact that a mechanical system has some kind of memory is not strange at all. For instance,
a pulse of light that impinges on a beam splitter induces a polarization in the
active part (usually a thin layer of metal) of the beam splitter~\cite{BORN64}. Assuming a linear response (as is
usually done in classical electrodynamics), we have ${\bf P}({\bf r},t)=\chi({\bf r},t)\ast {\bf E}(r,t)$ where ``$\ast$'' is a shorthand
for the convolution. If the susceptibility $\chi({\bf r},t)$ has a nontrivial time dependence (as in the Lorentz model~\cite{BORN64}
for instance), the polarization will exhibit ``memory'' effects and will ``learn'' from subsequent pulses.
DLMs mimic this behavior in the most simple manner (see the convolution in Eq.~(\ref{HYP2})), on an event-by-event basis.

\subsection{Mach-Zehnder interferometer}
\label{MZI}

\begin{figure}[t]
\begin{center}
\includegraphics[width=8.5cm]{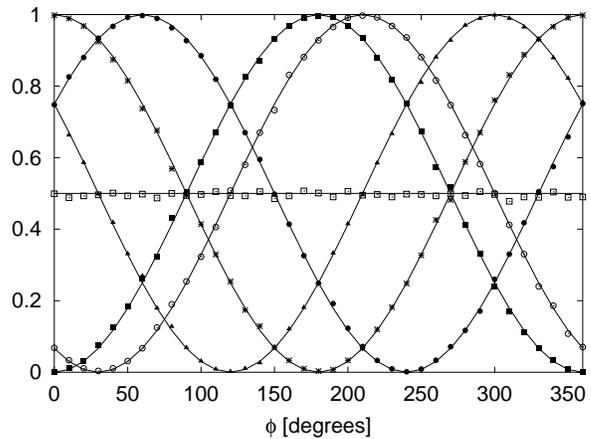}
\caption{
Simulation results for the \DLM-network shown in Fig.~\ref{onemzi}.
Input channel 0 receives $(\cos\psi_0,\sin\psi_0)$ with probability one.
A uniform random number in the range $[0,360]$ is used to choose the angle $\psi_0$.
Input channel 1 receives no events. 
Each data point represents 10000 events ($N_0+N_1=N_2+N_3=10000$).
Initially the rotation angle $\phi_0=0$ and after each set of 10000 events, $\phi_0$
is increased by $10^\circ$.
Markers give the simulation results for the normalized intensities
as a function of $\phi=\phi_0-\phi_1$.
Open squares: $N_0/(N_0+N_1)$;
Solid squares: $N_2/(N_2+N_3)$ for $\phi_1=0$;
Open circles: $N_2/(N_2+N_3)$ for $\phi_1=30^\circ$;
Bullets: $N_2/(N_2+N_3)$ for $\phi_1=240^\circ$;
Asterisks: $N_3/(N_2+N_3)$ for  $\phi_1=0$;
Solid triangles: $N_3/(N_2+N_3)$ for $\phi_1=300^\circ$.
Lines represent the results of quantum theory~\cite{QuantumTheory}.
}
\label{one-mz}
\label{figmz}
\end{center}
\end{figure}

\begin{figure*}[t]
\begin{center}
\includegraphics[width=16cm]{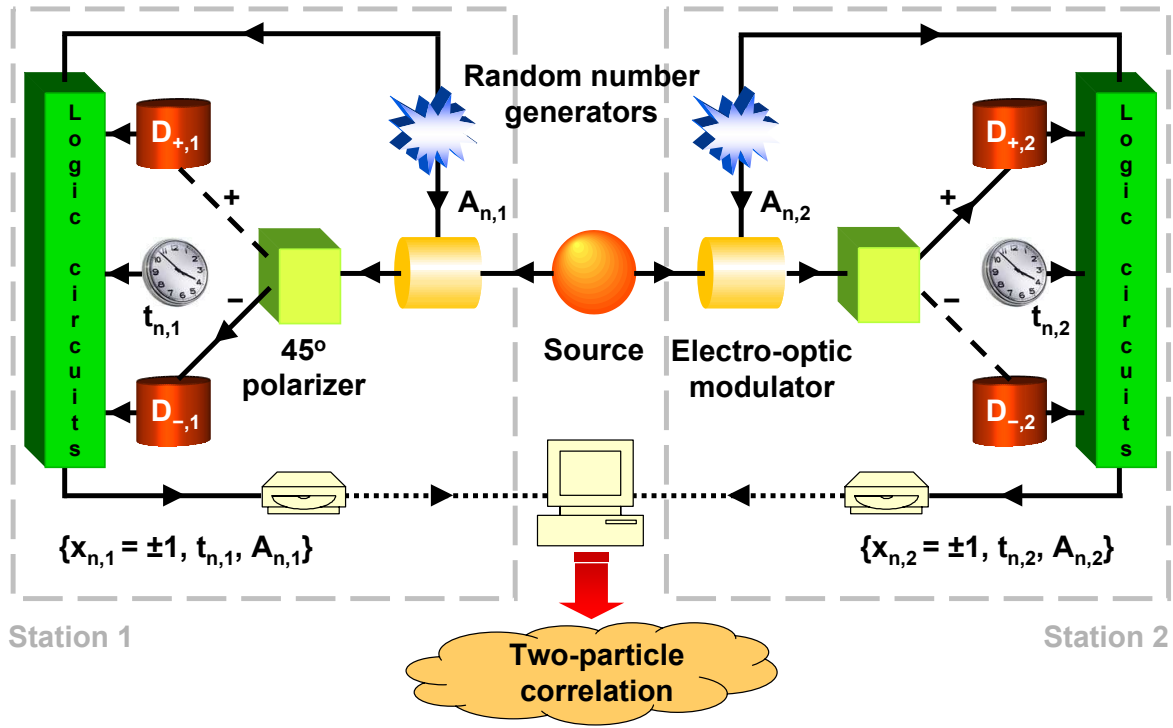}
\caption{(color online) Schematic diagram of an EPRB experiment with photons~\cite{WEIH98}.
}
\label{fig1a}
\end{center}
\end{figure*}

Using the \DLM\ of Fig.~\ref{one-bs} as a module that simulates a beam splitter,
we build the Mach-Zehnder interferometer by connecting
two \DLMS, as shown in Fig.~\ref{onemzi}.
The length of each path from the first to the second beam splitter
is made variable, as indicated by the controls on the horizontal lines.
The thin, $45^\circ$-tilted lines act as perfect mirrors.

In quantum theory, the presence of photons in
the input modes 0 or 1 of the interferometer is represented
by the probability amplitudes ($a_0,a_1)$~\cite{BAYM74}.
The amplitudes to observe a
photon in the output modes 0 and 1 of
the Mach-Zehnder interferometer (see Fig.~\ref{onemzi})
are given by
\begin{eqnarray}
\left(
\begin{array}{c}
b_2\\
b_3
\end{array}
\right)
=
\left(
\begin{array}{cc}
1&i\\
i&1
\end{array}
\right)
\left(
\begin{array}{cc}
e^{i\phi_0}&0\\
0&e^{i\phi_1}
\end{array}
\right)
\left(
\begin{array}{c}
b_0\\
b_1
\end{array}
\right)
,
\label{MZ1}
\end{eqnarray}
where $b_0$ and $b_1$ are given by Eq.~(\ref{BS3}).
In Eq.~(\ref{MZ1}), the entries
$e^{i\phi_j}$ for $j=0,1$ implement the phase shifts that result
from the time delays on the corresponding path
(including the phase shifts due to the presence of the perfect mirrors).

\subsection{Simulation results}

The snapshot in Fig.~\ref{onemzi} is taken after
$N=3030$ particles have been generated by the source.
The numbers in the various corresponding fields
clearly show that even after a modest number of
events, this event-by-event simulation reproduces
the quantum mechanical probabilities.
Of course, this single snapshot is not a proof that
the event-by-event simulation also works for
other choices of the time delays.
More extensive simulations, an example of a set of results being shown in Fig.~\ref{one-mz},
demonstrate that \DLM-networks accurately reproduce the probabilities
of quantum theory for these single-photon experiments~\cite{RAED05b,RAED05c,RAED05d,MICH05,RAED06z,MICH06z}.

\section{EPRB experiments}
\label{EPRBexperiment}

In Fig.~\ref{fig1a}, we show a schematic diagram of an EPRB experiment
with photons (see also Fig.~2 in~\cite{WEIH98}).
The source emits pairs of photons.
Each photon of a pair propagates to an observation station
in which it is manipulated and detected.
The two stations are separated spatially and temporally~\cite{WEIH98}.
This arrangement prevents the observation at
station 1 (2) to have a causal effect on the
data registered at station $2$ (1)~\cite{WEIH98}.
As the photon arrives at station $i=1,2$, it passes through an electro-optic
modulator that rotates the polarization of the photon by an angle depending
on the voltage applied to the modulator.
These voltages are controlled by two independent binary random number generators.
As the photon leaves the polarizer, it generates a signal in one of the
two detectors.
The station's clock assigns a time-tag to each generated signal.
Effectively, this procedure discretizes time in intervals of a width that is
determined by the time-tag resolution $\tau$~\cite{WEIH98}.
In the experiment, the firing of a detector is regarded as an event.

As we wish to demonstrate
that it is possible to reproduce the results of quantum theory (which implicitly assumes idealized conditions)
for the EPRB gedanken experiment by an event-based simulation algorithm,
it would be logically inconsistent to ``recover'' the results of the former
by simulating nonideal experiments.
Therefore, we consider ideal experiments only,
meaning that we assume that detectors operate with 100\% efficiency,
clocks remain synchronized forever, the ``fair sampling'' assumption is satisfied~\cite{ADEN07}, and so on.
We assume that the two stations are separated spatially and temporally
such that the manipulation and observation at station 1 (2) cannot have a causal effect on the
data registered at station $2$ (1).
Furthermore, to realize the EPRB gedanken experiment on the computer,
we assume that the orientation of each electro-optic modulator
can be changed at will, at any time.
Although these conditions are very difficult to satisfy in real experiments,
they are trivially realized in computer experiments.

In general, on logical grounds (without counterfactual reasoning),
it is impossible to make a statement about
the directions of the polarization of particles emitted by
the source unless we have performed an experiment to determine these directions.
However, in a computer experiment we have perfect control and we
can select any direction that we like.
Conceptually, there are two extreme cases.
In the first case, we assume that we know nothing about the direction
of the polarization. We mimic this situation by using
pseudo-random numbers to select the initial polarization.
This is the case that is typical for a real EPRB experiment.
In the second case, we assume that we know that
the polarizations are fixed (but are not necessarily the same),
mimicking a source that emits polarized photons.
A simulation algorithm that aims to reproduce all the results
of quantum theory should be able
to reproduce all these results for both cases
{\sl without any change to the simulation algorithm except for the part that
simulates the source}~\cite{RAED06c,RAED07a,RAED07c}.

In the experiment, the firing of a detector is regarded as an event.
At the $n$th event, the data recorded on a hard disk at station $i=1,2$
consists of $x_{n,i}=\pm 1$, specifying which of the two detectors fired,
the time tag $t_{n,i}$ indicating the time at which a detector fired,
and the two-dimensional unit vector ${\bf a}_{n,i}$ that represents the rotation
of the polarization by the electro-optic polarizer.
Hence, the set of data collected at station $i=1,2$ during a run of $N$ events
may be written as
\begin{eqnarray}
\label{Ups}
\Upsilon_i=\left\{ {x_{n,i} =\pm 1,t_{n,i},{\bf a}_{n,i} \vert n =1,\ldots ,N } \right\}
.
\end{eqnarray}
In the (computer) experiment, the data $\{\Upsilon_1,\Upsilon_2\}$ may be analyzed
long after the data has been collected~\cite{WEIH98}.
Coincidences are identified by comparing the time differences
$\{ t_{n,1}-t_{n,2} \vert n =1,\ldots ,N \}$ with a time window $W$~\cite{WEIH98}. 
Introducing the symbol $\sum'$ to indicate that the sum
has to be taken over all events that satisfy
$\mathbf{a}_i=\mathbf{a}_{n,i}$ for $i=1,2$,
for each pair of directions $\mathbf{a}_1$ and $\mathbf{a}_2$ of the electro-optic modulators,
the number of coincidences $C_{xy}\equiv C_{xy}(\mathbf{a}_1,\mathbf{a}_2)$ between detectors $D_{x,1}$ ($x =\pm 1$) at station
1 and detectors $D_{y,2}$ ($y =\pm1 $) at station 2 is given by
\begin{eqnarray}
\label{Cxy}
C_{xy}&=&\sumprime_{n=1}^N\delta_{x,x_{n ,1}} \delta_{y,x_{n ,2}}
\Theta(W-\vert t_{n,1} -t_{n ,2}\vert)
,
\end{eqnarray}
where $\Theta (t)$ is the Heaviside step function.
We emphasize that we count
all events that, according to the same criterion as the one employed in experiment,
correspond to the detection of pairs.
The average single-particle counts are defined by
\begin{eqnarray}
\label{Ex}
E_1(\mathbf{a}_1,\mathbf{a}_2)&=&
\frac{\sum_{x,y=\pm1} xC_{xy}}{\sum_{x,y=\pm1} C_{xy}}
,
\nonumber \\
\noalign{and}
E_2(\mathbf{a}_1,\mathbf{a}_2)&=&\frac{\sum_{x,y=\pm1} yC_{xy}}{\sum_{x,y=\pm1} C_{xy}}
,
\end{eqnarray}
where the denominator is the sum of all coincidences.

According to standard terminology, the correlation between $x=\pm1$ and $y=\pm1$ events is defined by~\cite{GRIM95}
\begin{widetext}
\begin{eqnarray}
\label{rhoxy}
\rho(\mathbf{a}_1,\mathbf{a}_2)&=&
\frac{
\frac{\sum_{x,y} xyC_{xy}}{\sum_{x,y} C_{xy}}
-\frac{\sum_{x,y} xC_{xy}}{\sum_{x,y} C_{xy}}
\frac{\sum_{x,y} yC_{xy}}{\sum_{x,y} C_{xy}}
}{
\sqrt{
\left(
\frac{ \sum_{x,y} x^2C_{xy} }{\sum_{x,y} C_{xy}}-
(\frac{\sum_{x,y} xC_{xy} }{\sum_{x,y} C_{xy} })^2
\right)
\left(
\frac{\sum_{x,y} y^2C_{xy}}{\sum_{x,y} C_{xy}}-
(\frac{(\sum_{x,y} yC_{xy}}{\sum_{x,y} C_{xy}})^2
\right)
}
}
.
\end{eqnarray}
\end{widetext}

The correlation $\rho(\mathbf{a}_1,\mathbf{a}_2)$
is $+1$ ($-1$) in the case that $x=y$ ($x=-y$) with certainty.
If the values of $x$ and $y$ are independent, the correlation $\rho(\mathbf{a}_1,\mathbf{a}_2)$
is zero.  Note that in general, the converse is not necessarily true but in the special case
of dichotomic variables $x$ and $y$, the converse is true~\cite{RAED07b}.

In the case of dichotomic variables $x$ and $y$, the correlation $\rho(\mathbf{a}_1,\mathbf{a}_2)$
is entirely determined by the average single-particle counts Eq.~(\ref{Ex})
and the two-particle average
\begin{eqnarray}
\label{Exy}
E(\mathbf{a}_1,\mathbf{a}_2)&=&
\frac{\sum_{x,y} xyC_{xy}}{\sum_{x,y} C_{xy}}
\nonumber \\
&=&\frac{C_{++}+C_{--}-C_{+-}-C_{-+}}{C_{++}+C_{--}+C_{+-}+C_{-+}}
.
\end{eqnarray}
For later use, it is expedient to introduce the function
\begin{equation}
\label{Sab}
S(\mathbf{a},\mathbf{b},\mathbf{c},\mathbf{d})=
E(\mathbf{a},\mathbf{c})-E(\mathbf{a},\mathbf{d})
+
E(\mathbf{b},\mathbf{c})+E(\mathbf{b},\mathbf{d})
,
\end{equation}
and its maximum
\begin{eqnarray}
\label{Smax}
S_{max}&\equiv&\max_{\mathbf{a},\mathbf{b},\mathbf{c},\mathbf{d}} S(\mathbf{a},\mathbf{b},\mathbf{c},\mathbf{d})
.
\end{eqnarray}

In general, the values for the average
single-particle counts $E_1(\mathbf{a}_1,\mathbf{a}_2)$ and $E_2(\mathbf{a}_1,\mathbf{a}_2)$
the coincidences $C_{xy}(\mathbf{a}_1,\mathbf{a}_2)$,
the two-particle averages $E(\mathbf{a}_1,\mathbf{a}_2)$,
$S(\mathbf{a},\mathbf{b},\mathbf{c},\mathbf{d})$, and $S_{max}$
not only depend on the directions $\mathbf{a}_1$ and $\mathbf{a}_2$ but also
on the time-tag resolution $\tau$
and the time window $W$ used to identify the coincidences.

\begin{figure}[t]
\begin{center}
\includegraphics[width=8cm]{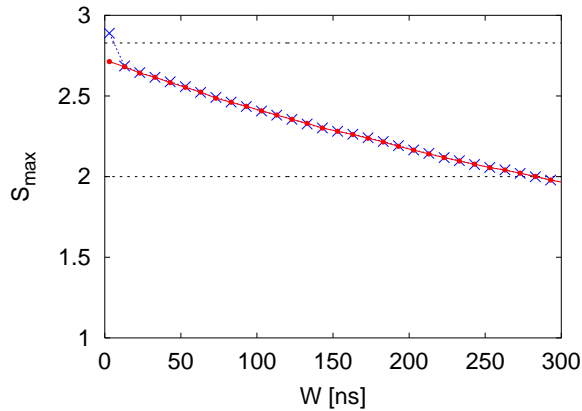}
\caption{(color online) $S_{max}$ as a function of the time window $W$,
computed from the data sets contained in the archives
Alice.zip and Bob.zip that can be downloaded from Ref.~\onlinecite{WEIHdownload}.
Bullets (red): Data obtained by
using the relative time shift $\Delta=4$ ns that maximizes the
number of coincidences.
Crosses (blue): Raw data ($\Delta=0$).
Dashed line at $2\sqrt 2 $: $S_{max}$ if the system is described by quantum theory (see Section~\ref{Quantumtheory}).
Dashed line at $2$: $S_{max}$ if the system is described by the
class of models introduced by Bell~\cite{BELL93}.
}
\label{exp1}
\end{center}
\end{figure}

\subsection{Analysis of real experimental data}
\label{IIG}

We illustrate the procedure of data analysis and
the importance of the choice of the time window
$W$ by analyzing a data set (the archives
Alice.zip and Bob.zip) of an EPRB experiment with photons
that is publicly available~\cite{WEIHdownload}.

In the real experiment, the number of events detected at station 1 is unlikely
to be the same as the number of events detected at station 2.
In fact, the data sets of Ref.~\onlinecite{WEIHdownload} show that
station 1 (Alice.zip) recorded 388455 events while
station 2 (Bob.zip) recorded  302271 events.
Furthermore, in the real EPRB experiment, there may be an
unknown shift $\Delta$ (assumed to be constant during the experiment)
between the times $t_{n,1}$ gathered at station 1 and
the times $t_{n,2}$ recorded at station 2.
Therefore, there is some extra ambiguity in
matching the data of station 1 to the data of station 2.

\begin{figure}[t]
\begin{center}
\includegraphics[width=8cm]{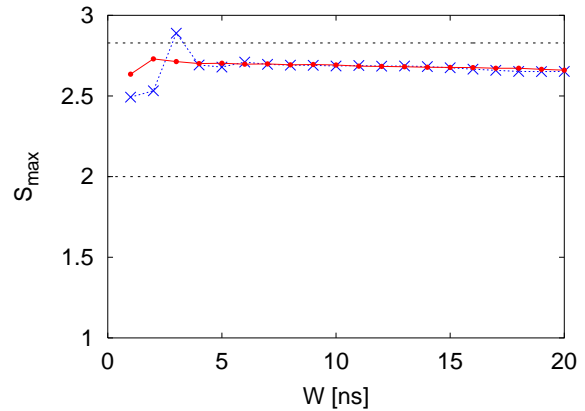}
\caption{(color online) Same as Fig.~\ref{exp1} except for the range of $W$.
Bullets (red): Data obtained by
using the relative time shift $\Delta=4$ ns that maximizes the
number of coincidences.
The maximum value of $S_{max}\approx2.73$ is found at $W=2$ ns.
Crosses (blue): Raw data $\Delta=0$.
The maximum value of $S_{max}\approx2.89$ is found at $W=3$ ns.
}
\label{exp2}
\end{center}
\end{figure}

A simple data processing procedure that resolves this
ambiguity consists of two steps~\cite{WEIH00}.
First, we make a histogram of the time differences
$t_{n,1}-t_{m,2}$ with a small but reasonable resolution
(we used $0.5$ ns).
Then, we fix the value of the time-shift $\Delta$
by searching for the time difference for which
the histogram reaches its maximum, that is we maximize
the number of coincidences by a suitable choice of $\Delta$.
For the case at hand, we find $\Delta=4$ ns.
Finally, we compute the coincidences,
the two-particle average, and $S_{max}$ using
the expressions given earlier.
The average times between two detection events is $2.5$ ms and $3.3$ ms
for Alice and Bob, respectively.
The number of coincidences (with double counts removed) is
13975 and 2899 for ($\Delta=4$ ns, $W=2$ ns) and
($\Delta=0$ , $W=3$ ns) respectively.

In Figs.~\ref{exp1} and \ref{exp2} we present the results
for $S_{max}$ as a function of the time window $W$.
First, it is clear that $S_{max}$ decreases significantly
as $W$ increases but it is also clear that as $W\rightarrow0$,
$S_{max}$ is not very sensitive to the choice of $W$~\cite{WEIH00}.
Second, the procedure of maximizing the coincidence count
by varying $\Delta$ reduces the maximum value of $S_{max}$ from
a value 2.89 that considerably exceeds the maximum
for the quantum system ($2\sqrt{2}$, see Section~\ref{Quantumtheory}) to a value 2.73 that
violates the Bell inequality ($S_{max}\le2$, see Ref.~\onlinecite{BELL93}) and is less than
the maximum for the quantum system.

The fact that the ``uncorrected'' data ($\Delta=0$) violate the rigorous bound for the quantum system
should not been taken as evidence that quantum theory is ``wrong'':
It merely indicates that the way in which the data of the two stations
has been grouped in two-particle events is not optimal.
There is no reason why a correlation between
similar but otherwise unrelated data should be described by quantum theory.

\begin{figure}[t]
\begin{center}
\includegraphics[width=8.5cm]{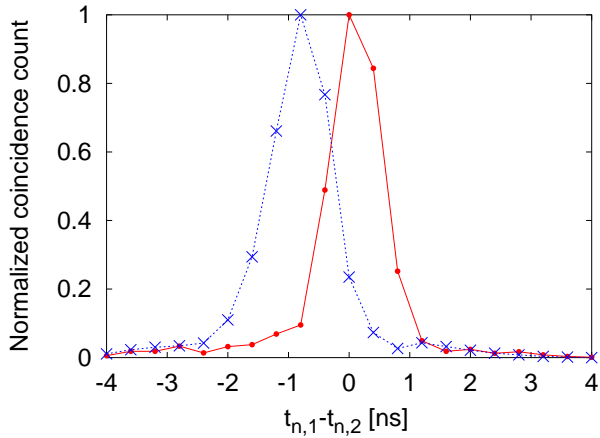}
\caption{(color online) Normalized coincidence counts as a function of time tag difference $t_{n,1}-t_{n,2}$,
computed from the data sets contained in the archives
Alice.zip and Bob.zip~\cite{WEIHdownload}, using the relative time shift $\Delta=4$ ns that maximizes the
number of coincidences.
Bullets (red): $\theta_1=0$ and $\theta_2=\pi/8$;
Crosses (blue): $\theta_1=0$ and $\theta_2=3\pi/8$.
}
\label{fort.7}
\end{center}
\end{figure}

Finally, we use the experimental data to show that the time delays depend on the
orientation of the polarizer. To this end, we
select all coincidences between $D_{+,1}$ and $D_{+,2}$ (see Fig.~\ref{fig1a}) and
make a histogram of the coincidence counts as a function of the time-tag difference,
for fixed orientation $\theta_1=0$ and the two orientations $\theta_2=\pi/8,3\pi/8$ (other
combinations give similar results).
The results of this analysis are shown in Fig.~\ref{fort.7}.
The maximum of the distribution shifts by approximately 1 ns as the polarizer at station 2 is
rotated by $\pi/4$, a demonstration that the time-tag data is sensitive to the orientation
of the polarizer at station 2. A similar distribution of time-delays (of about the same width) was also observed in a much older
experimental realization of the EPRB experiment~\cite{KOCH67}.

According to Maxwell's equation, the birefringent properties of the optically anisotropic materials that are
used to fabricate the optical elements (polarizers and electro-optic modulators),
cause plane waves with different polarization to propagate
with different phase velocity~\cite{BORN64}, suggesting a possible mechanism for the time delays observed in experiments.
As light is supposed to consist of non-interacting photons,
this suggests, but does not prove, that individual photons
experience a time delay as they pass through the electro-optic modulators or polarizers.
Of course, strictly speaking, we cannot derive the time delay from classical electrodynamics:
The concept of a photon has no place in Maxwell's theory.
A more detailed understanding of the time delay mechanism first
requires dedicated, single-photon retardation measurements for these specific optical elements.

\subsection{Role of the coincidence window $W$}

The crucial point is that in any real EPR-type experiment,
it is necessary to have an operational procedure to decide
if the two detection events correspond to the observation of
one two-particle system or to the observation of two single-particle systems.
In standard ``hidden variable'' treatments of the EPR gedanken experiment~\cite{BELL93},
the operational definition of ``observation of a single two-particle system'' is missing.
In EPRB-type experiments, this decision is taken on the basis of coincidence in time~\cite{KOCH67,CLAU74,WEIH98}.

Our analysis of the experimental data shows beyond doubt
that a model which aims to describe real EPRB experiments
should include the time window $W$ and that the interesting regime
is $W\rightarrow0$, not $W\rightarrow\infty$ as is assumed
in all textbook treatments of the EPRB experiment.
Indeed, in quantum mechanics textbooks it is standard to assume that an EPRB experiment
measures the correlation~\cite{BELL93}
\begin{eqnarray}
\label{CxyBell}
C_{xy}^{(\infty)}&=&\sumprime_{n=1}^N\delta_{x,x_{n ,1}} \delta_{y,x_{n ,2}}
,
\end{eqnarray}
which we obtain from Eq.~(\ref{Cxy}) by taking the limit $W\rightarrow\infty$.
Although this limit defines a valid theoretical model, there is no reason
why this model should have any bearing on the real experiments, in particular
because experiments pay considerable attention to the choice of $W$.
A rational argument that might justify taking this limit is
the hypothesis that for ideal experiments, the value of $W$ should
not matter. However, in experiments a lot of effort is made to reduce (not increase) $W$~\cite{WEIH98,WEIH00}.

As we will see later, using our model it is relatively easy to reproduce the experimental facts
and the results of quantum theory if we consider the limit $W\rightarrow0$.
Furthermore, keeping $W$ arbitrary does not render the mathematics more complicated
so there really is no point of studying the simplified model defined by Eq.~(\ref{CxyBell}):
We may always consider the limiting case $W\rightarrow\infty$ afterwards.

\subsection{Quantum theory}
\label{Quantumtheory}
According to the axioms of quantum theory~\cite{BALL03},
repeated measurements on the two-spin system described by the density matrix $\rho$
yield statistical estimates
for the single-spin expectation values
\begin{equation}
\label{Ei}
\widetilde E_1(\mathbf{a})=\langle \mathbf{\sigma}_1\cdot\mathbf{a} \rangle
\quad,\quad
\widetilde E_2(\mathbf{b})=\langle \mathbf{\sigma}_2\cdot\mathbf{b} \rangle
,
\end{equation}
and the two-spin expectation value
\begin{eqnarray}
\label{Eab}
\widetilde E(\mathbf{a},\mathbf{b})&=&
\langle \mathbf{\sigma}_1\cdot\mathbf{a}\; \mathbf{\sigma}_2\cdot\mathbf{b} \rangle
,
\end{eqnarray}
where $\sigma_i=(\sigma_i^x ,\sigma_i^y ,\sigma_i^z )$
are the Pauli spin-1/2 matrices describing the spin of particle $i=1,2$~\cite{BALL03},
and $\mathbf{a}$ and $\mathbf{b}$ are unit vectors.
We have introduced the tilde to distinguish the quantum theoretical results from the results
obtained from the data sets $\{\Upsilon_1,\Upsilon_2\}$.
The state of a quantum system of two $S=1/2$ objects is completely determined
if we know the expectation values $\widetilde E_1(\mathbf{a})$,
$\widetilde E_2(\mathbf{b})$, and $\widetilde E(\mathbf{a},\mathbf{b})$.

It can be shown that $|\widetilde S(\mathbf{a},\mathbf{b},\mathbf{c},\mathbf{d})|\le2\sqrt{2}$~\cite{CIRE80},
independent of the choice of $\rho$.
If the density matrix $\rho=\rho_1\otimes\rho_2$ factorizes
(here $\rho_i$ is the $2\times2$ density matrix of spin $i$),
then it is easy to prove that $|\widetilde S(\mathbf{a},\mathbf{b},\mathbf{c},\mathbf{d})|\le2$.
In other words, if
$\max_{\mathbf{a},\mathbf{b},\mathbf{c},\mathbf{d}} \widetilde S(\mathbf{a},\mathbf{b},\mathbf{c},\mathbf{d})>2$,
then $\rho\not=\rho_1\otimes\rho_2$, and the quantum system is in an entangled state.
Specializing to the case of the photon polarization,
the unit vectors $\mathbf{a}$, $\mathbf{b}$, $\mathbf{c}$, and $\mathbf{d}$
lie in the same plane and we may use the angles
$\alpha$, $\alpha'$, $\beta$, and $\beta'$
to specify their direction.

The quantum theoretical description of the EPRB experiment
assumes that the system is represented by the singlet state
$|\Psi\rangle=\left(| H\rangle_1 | V \rangle_2 -| V \rangle _1 |H \rangle_2\right)/\sqrt 2 $
of two spin-1/2 particles,
where $H$ and $V$ denote the horizontal
and vertical polarization and the subscripts refer to photon 1 and 2, respectively.
For the singlet state $\rho=|\Psi\rangle\langle\Psi|$, 
\begin{eqnarray}
\label{E1}
\widetilde E_1(\alpha)&=&\widetilde E_2(\beta)=0,
\\
\label{eq2}
\widetilde E(\alpha,\beta)&=&-\cos 2(\alpha -\beta)
,
\end{eqnarray}
for which $\max_{\alpha,\alpha',\beta,\beta'} \widetilde S(\alpha,\alpha',\beta,\beta')=2\sqrt{2}$,
confirming that the singlet is a quantum state with maximal entanglement.

Analysis of the experimental data according
to the procedure sketched earlier~\cite{FREE72,ASPE82b,TAPS94,TITT98,WEIH98,ROWE01,FATA04},
yields results that are in good agreement with
$\widetilde E_1(\alpha)=\widetilde E_2(\beta)=0$ and $\widetilde E(\alpha,\beta)=-\cos 2(\alpha -\beta)$,
leading to the conclusion that in a quantum theoretical description,
the density matrix does not factorize, in spite of the fact that the photons are
spatially and temporally separated and do not interact.

\subsection{Classical simulation model}
\label{SimulationModel}

A concrete simulation model of the EPRB experiment sketched in Fig.~\ref{fig1a} requires
a specification of the information carried by the particles,
of the algorithm that simulates the source and
the observation stations, and of the procedure to analyze the data.
In the following, we describe a slightly modified version
of the algorithm proposed in Ref.~\cite{RAED06c}, tailored
to the case of photon polarization.

{\bf Source and particles:}
The source emits particles that carry a 
vector ${\bf S}_{n,i}=(\cos (\xi_{n}+(i-1)\pi/2) ,\sin (\xi_{n}+(i-1)\pi/2)$,
representing the polarization of the photons
that travel to station $i=1$ and station $i=2$, respectively.
Note that ${\bf S}_{n,1}\cdot {\bf S}_{n,2}=0$, indicating that
the two particles have orthogonal polarizations.
The ``polarization state'' of a particle is completely characterized by $\xi _{n}$,
which is distributed uniformly over the whole interval $[0,2\pi[$.
For this purpose, to mimic the apparent unpredictability of the experimental data,
we use uniform random numbers.
However, from the description of the algorithm, it will be clear that
the use of random numbers is not essential.
Simple counters that sample
the intervals $[0,2\pi[$ in a systematic, but uniform, manner might be employed
as well.

{\bf Observation station:}
The electro-optic modulator in station $i$ rotates ${\bf S}_{n,i}$ by an angle $\gamma _{n ,i}$, that is
${\bf a}_{n,i}=(\cos\gamma _{n ,i},\sin\gamma _{n ,i})$.
The number $M$ of different rotation angles is
chosen prior to the data collection (in the experiment of Weihs \textit{et al.}, $M=2$~\cite{WEIH98}).
We use $2M$ random numbers to fill the arrays $(\alpha _1 ,...,\alpha _M)$ and
$(\beta _1 ,...,\beta _M)$.
During the measurement process we use two uniform random numbers $1\le m,m'\le M$
to select the rotation angles $\gamma _{n,1} =\alpha _m$ and $\gamma _{n,2} =\beta _{{m}'}$.
The electro-optic modulator then rotates ${\bf S}_{n ,i} =(\cos (\xi _{n} +(i-1)\pi/2),\sin (\xi _{n}+(i-1)\pi/2 )$
by $\gamma _{n,i}$, yielding ${\bf S}_{n,i}=(\cos (\xi _{n} -\gamma_{n,i}+(i-1)\pi/2 ),\sin (\xi _{n} -\gamma_{n,i}+(i-1)\pi/2 )$.

The polarizer at station $i$ projects the rotated vector
onto its $x$-axis: ${\bf S}_{n ,i} \cdot \hat {\bf x}_i =\cos (\xi _{n } -\gamma _{n ,i}+(i-1)\pi/2 )$,
where $\hat {\bf  x}_i $ denotes the unit vector along the $x$-axis of the polarizer.
For the polarizing beam splitter, we consider a simple model: If $\cos ^2(\xi _{n } -\gamma _{n ,i}+(i-1)\pi/2 )>1/2$
the particle causes $D_{+1,i}$ to fire, otherwise $D_{-1,i}$ fires.
Thus, the detection of the particles generates the data
$x_{n ,i} =\mathop{\hbox{sign}}(\cos 2(\xi _{n } -\gamma _{n ,i} +(i-1)\pi/2))$.

{\bf Time-tag model:}
To assign a time-tag to each event,
we assume that as a particle passes through the detection system, it may experience a time delay.
In our model, the time delay ${t}_{n ,i} $ for a particle
is assumed to be distributed uniformly over the interval $[t_{0}, t_{0}+T]$,
an assumption that is not in conflict with available data~\cite{WEIH00}.
In practice, we use uniform random numbers
to generate ${t}_{n ,i}$.
As in the case of the angles $\xi_{n}$, the random choice of ${t}_{n ,i}$
is merely convenient, not essential.
From Eq.(\ref{Cxy}), it follows that only differences of time delays matter.
Hence, we may put $t_0=0$.
The time-tag for the event $n$ is then $t_{n,i}\in[0,T]$.

There are not many options to make a reasonable choice for $T$.
Assuming that the particle ``knows'' its own direction
and that of the polarizer only, we can construct one
number that depends on the relative angle: ${\bf S}_{n ,i} \cdot \hat {\bf x}_i$.
Thus, $T=T(\xi _{n } -\gamma _{n ,i} )$ depends on $\xi _{n } -\gamma _{n ,i} $ only.
Furthermore, consistency with classical electrodynamics requires that
functions that depend on the polarization have period $\pi$~\cite{BORN64}.
Thus, we must have $T(\xi _{n } -\gamma _{n ,i} +(i-1)\pi/2)=F( ({\bf S}_{n ,i} \cdot \hat {\bf x}_i)^2)$.
We already used $\cos 2(\xi _{n } -\gamma _{n ,i}+(i-1)\pi/2 )$ to determine whether the particle generates
a $+1$ or $-1$ signal.
By trial and error, we found that $T(\xi _{n } -\theta_1 )=T_0 F(|\sin 2(\xi _{n } -\theta_1)|)=T_0|\sin 2(\xi _{n } -\theta_1)|^d$
yields useful results~\cite{RAED06c,RAED07b,RAED07a,RAED07c,RAED07d}.
Here, $T_0 =\max_\theta T(\theta)$ is the maximum time delay
and defines the unit of time, used in the simulation and $d$ is a free parameter of the model.
In our numerical work, we set $T_0=1$.

{\bf Data analysis:}
For fixed $N$ and $M$, the algorithm generates the
data sets $\Upsilon_i$
just as experiment does~\cite{WEIH98}.
In order to count the coincidences, we choose a
time-tag resolution $0<\tau<T_0 $ and a coincidence window $\tau\le W$.
We set the correlation counts $C_{xy} (\alpha _m ,\beta _{m'} )$ to zero for all $x,y=\pm 1$
and $m,m'=1,...,M$.
We compute the discretized time tags $k_{n ,i} =\lceil t_{n ,i}/ \tau\rceil $
for all events in both data sets.
Here $\lceil{x}\rceil$ denotes the smallest integer that is larger or equal to $x$, that is
$\lceil{x}\rceil-1<x\le\lceil{x}\rceil$.
According to the procedure adopted in the experiment~\cite{WEIH98},
an entangled photon pair is observed if and only if
$\left| {k_{n,1} -k_{n,2} } \right|<k=\lceil{W/\tau}\rceil$.
Thus, if $\left| {k_{n,1} -k_{n,2} } \right|<k$,
we increment the count $C_{x_{n,1},x_{n,2}} (\alpha _m ,\beta _{m'} )$.

\begin{figure}[t]
\begin{center}
\mbox{
\includegraphics[width=8.5cm]{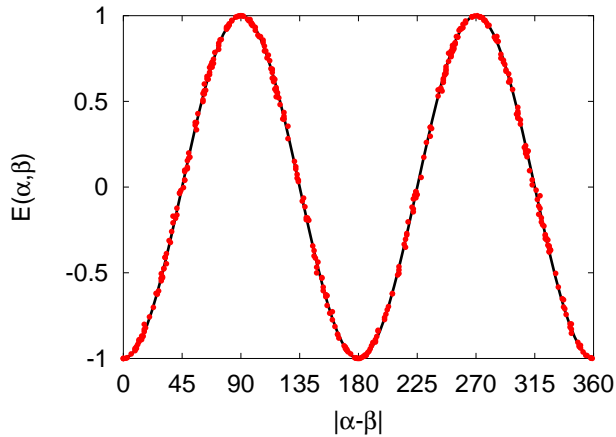}
}
\caption{(color online)
Comparison between computer simulation data (red bullets) and quantum theory
(black solid line) for the two-particle correlation $E(\alpha,\beta)$.
}
\label{fig2}
\label{fig4}
\end{center}
\end{figure}

\subsection{Simulation results}

The simulation proceeds in the same way as the experiment, that is we first
collect the data sets $\{\Upsilon_1,\Upsilon_2\}$, and then compute the coincidences
Eq.~(\ref{Cxy}) and the correlation Eq.~(\ref{Exy}).
The simulation results for the coincidences $C_{xy} (\alpha,\beta )$ depend on the
time-tag resolution $\tau$, the time window $W$ and the number of events $N$,
just as in real experiments~\cite{FREE72,ASPE82a,ASPE82b,TAPS94,TITT98,WEIH98,ROWE01,FATA04}.

Figure~\ref{fig2} shows simulation data for $E(\alpha ,\beta )$
as obtained for $N=10^6$ and $W=\tau=0.00025T_0$.
In the experiment, for each event, the random numbers $A_{n,i}=1,\ldots,M$ select one out of four pairs
$\{ (\alpha_i,\beta_j)|i,j=1,M\}$, where
the angles $\alpha_i$ and $\beta_i$ are fixed before the data is recorded.
The data shown has been obtained by allowing for $M=20$ different angles per station.
Hence, forty random numbers
from the interval [0,360[ were used to fill the arrays $(\alpha _1,\ldots,\alpha _M )$
and $(\beta _1 ,\ldots,\beta _M )$.
For each of the $N $ events, two different random number generators were used to select the
angles $\alpha _m $ and $\beta _{{m}'} $. The statistical correlation between $m$ and
$m'$ was measured to be less than $10^{-6}$.

From Fig.~\ref{fig2}, it is clear that the simulation data for $E(\alpha,\beta )$ are
in excellent agreement with quantum theory.
Within the statistical noise, the simulation data (not shown)
for the single-spin expectation values also
reproduce the results of quantum theory.

Additional simulation results (not shown) demonstrate that the kind of models described
earlier are capable of reproducing all the results of quantum theory for
a system of two S=1/2 particles~\cite{RAED06c,RAED07b,RAED07a,RAED07c,RAED07d}.
Furthermore, for $W\rightarrow0$ and in the limit that the number of events goes
to infinity, one can prove rigorously that these simulation models
give the same expressions for the single- and two-particle averages
as those obtained from quantum theory~\cite{RAED06c,RAED07b,RAED07a,RAED07c,RAED07d}.

\subsection{Discussion}

Starting from the factual observation that experimental realizations of the EPRB experiment
produce the data $\{\Upsilon_1,\Upsilon_2\}$ (see Eq.~(\ref{Ups})) and that coincidence in time is
a key ingredient for the data analysis,
we have described a computer simulation model that satisfies Einstein's criterion of local causality
and, exactly reproduces the correlation $\widetilde E(\mathbf{a}_1,\mathbf{a}_2)=-\mathbf{a}_1\cdot\mathbf{a}_2$
that is characteristic for a quantum system in the singlet state.
Salient features of these models are that they
generate the data set Eq.~(\ref{Ups}) event-by-event,
use integer arithmetic and elementary mathematics to analyze the data,
do not rely on concepts of probability and quantum theory,
and provide a simple, rational and realistic picture of the mechanism that yields correlations such as Eq.~(\ref{eq2}).

We have shown that whether or not these simulation models produce quantum correlations
depends on the data analysis procedure that is performed (long) after the data has been collected:
In order to observe the correlations of the singlet state,
the resolution $\tau$ of the devices that generate the
time-tags and the time window $W$ should be made as small as possible.
Disregarding the time-tag data ($d=0$ or $W>T_0$) yields results that disagree with quantum theory
but agree with the models considered by Bell~\cite{BELL93}.
Our analysis of real experimental data and our simulation
results show that increasing the time window changes the nature of the
two-particle correlations~\cite{RAED06c,RAED07b,RAED07a,RAED07c,RAED07d}.

According to the folklore about Bell's theorem, a procedure
such as the one that we described should not exist.
Bell's theorem states that any local, hidden variable
model will produce results that are in conflict with the quantum theory of a system of two $S=1/2$ particles~\cite{BELL93}.
However, it is often overlooked that this statement can be proven for a (very) restricted class of probabilistic models only.
Indeed, minor modifications to the original model of Bell lead to the conclusion that there is no conflict~\cite{LARS04,SANT05,ZUKO06}.
In fact, Bell's theorem does not necessarily apply to the systems that we are interested in
as both simulation algorithms and actual data do not need to
satisfy the (hidden) conditions under which Bell's theorem hold~\cite{SICA99,HESS04,HESS05b}.

The apparent conflict between the fact that there exist event-based simulation models
that satisfy Einstein's criterion of local causality and reproduce
all the results of the quantum theory of a system of two $S=1/2$ particles
and the folklore about Bell's theorem, stating that such models are not supposed to exist
dissolves immediately if one recognizes that
Bell's extension of Einstein's concept of locality to the domain of probabilistic theories
relies on the hidden, fundamental assumption
that the absence of a causal influence implies logical independence~\cite{JAYN89}.
Indeed, in an attempt to extend Einstein's concept of a locally causal theory
to probabilistic theories, Bell implicitly assumed
that the absence of causal influence implies logical independence.
In general, this assumption prohibits the consistent application of probability theory and
leads to all kinds of logical paradoxes~\cite{TRIB69,JAYN03}.
However, if we limit our thinking to the domain of quantum physics,
the violation of the Bell inequalities by experimental data
should be taken as a strong signal that it is the correctness of this assumption
that one should question. Thus, we are left with two options:
\begin{itemize}
\item{One accepts the assumption that the absence of a causal influence implies logical independence
and lives with the logical paradoxes that this assumption creates.}
\item{One recognizes that logical independence and the absence of a causal influence are
different concepts~\cite{COX61,TRIB69,JAYN03} and one searches for rational explanations
of experimental facts that are logically consistent, as we did in our simulational
approach.}
\end{itemize}

\section{Conclusion}
\label{Conclusions}

The simulation models that I described in this talk do not
rely on concepts of probability theory or quantum theory: They are purely ontological models of quantum phenomena.
The salient features of these simulation models~\cite{RAED05b,RAED05c,RAED05d,MICH05,RAED06c,RAED07a,RAED07b,ZHAO07b} are that they
\begin{enumerate}
\item{generate, event-by-event, the same type of data as recorded in experiment,}
\item{analyze data according to the procedure used in experiment,}
\item{satisfy Einstein's criterion of local causality,}
\item{do not rely on any concept of quantum theory or probability theory,}
\item{reproduce the averages that we compute from quantum theory,}
\end{enumerate}
We may therefore conclude that this computational modeling approach opens new routes to ontological descriptions of
microscopic phenomena.

\medskip
\begin{acknowledgments}
I thank K. Michielsen for a critical reading of the manuscript.
\end{acknowledgments}

\bibliography{epr}

\begin{thebibliography}{49}
\expandafter\ifx\csname natexlab\endcsname\relax\def\natexlab#1{#1}\fi
\expandafter\ifx\csname bibnamefont\endcsname\relax
  \def\bibnamefont#1{#1}\fi
\expandafter\ifx\csname bibfnamefont\endcsname\relax
  \def\bibfnamefont#1{#1}\fi
\expandafter\ifx\csname citenamefont\endcsname\relax
  \def\citenamefont#1{#1}\fi
\expandafter\ifx\csname url\endcsname\relax
  \def\url#1{\texttt{#1}}\fi
\expandafter\ifx\csname urlprefix\endcsname\relax\def\urlprefix{URL }\fi
\providecommand{\bibinfo}[2]{#2}
\providecommand{\eprint}[2][]{\url{#2}}

\bibitem[{\citenamefont{Landau and Binder}(2000)}]{LAND00}
\bibinfo{author}{\bibfnamefont{D.~P.} \bibnamefont{Landau}} \bibnamefont{and}
  \bibinfo{author}{\bibfnamefont{K.}~\bibnamefont{Binder}},
  \emph{\bibinfo{title}{A Guide to Monte Carlo Simulation in Statistical
  Physics}} (\bibinfo{publisher}{Cambridge University Press},
  \bibinfo{address}{Cambridge}, \bibinfo{year}{2000}).

\bibitem[{\citenamefont{Bohm}(1951)}]{BOHM51}
\bibinfo{author}{\bibfnamefont{D.}~\bibnamefont{Bohm}},
  \emph{\bibinfo{title}{Quantum Theory}} (\bibinfo{publisher}{Prentice-Hall},
  \bibinfo{address}{New York}, \bibinfo{year}{1951}).

\bibitem[{\citenamefont{Home}(1997)}]{HOME97}
\bibinfo{author}{\bibfnamefont{D.}~\bibnamefont{Home}},
  \emph{\bibinfo{title}{Conceptual Foundations of Quantum Physics}}
  (\bibinfo{publisher}{Plenum Press}, \bibinfo{address}{New York},
  \bibinfo{year}{1997}).

\bibitem[{\citenamefont{Ballentine}(2003)}]{BALL03}
\bibinfo{author}{\bibfnamefont{L.~E.} \bibnamefont{Ballentine}},
  \emph{\bibinfo{title}{Quantum Mechanics: A Modern Development}}
  (\bibinfo{publisher}{World Scientific}, \bibinfo{address}{Singapore},
  \bibinfo{year}{2003}).

\bibitem[{\citenamefont{Tonomura}(1998)}]{TONO98}
\bibinfo{author}{\bibfnamefont{A.}~\bibnamefont{Tonomura}},
  \emph{\bibinfo{title}{The Quantum World Unveiled by Electron Waves}}
  (\bibinfo{publisher}{World Scientific}, \bibinfo{address}{Singapore},
  \bibinfo{year}{1998}).

\bibitem[{\citenamefont{Grangier et~al.}(1986)\citenamefont{Grangier, Roger,
  and Aspect}}]{GRAN86}
\bibinfo{author}{\bibfnamefont{P.}~\bibnamefont{Grangier}},
  \bibinfo{author}{\bibfnamefont{G.}~\bibnamefont{Roger}}, \bibnamefont{and}
  \bibinfo{author}{\bibfnamefont{A.}~\bibnamefont{Aspect}},
  \bibinfo{journal}{Europhys. Lett.} \textbf{\bibinfo{volume}{1}},
  \bibinfo{pages}{173} (\bibinfo{year}{1986}).

\bibitem[{\citenamefont{{De Raedt} et~al.}(2005{\natexlab{a}})\citenamefont{{De
  Raedt}, {De Raedt}, and Michielsen}}]{RAED05b}
\bibinfo{author}{\bibfnamefont{K.}~\bibnamefont{{De Raedt}}},
  \bibinfo{author}{\bibfnamefont{H.}~\bibnamefont{{De Raedt}}},
  \bibnamefont{and}
  \bibinfo{author}{\bibfnamefont{K.}~\bibnamefont{Michielsen}},
  \bibinfo{journal}{Comp. Phys. Comm.} \textbf{\bibinfo{volume}{171}},
  \bibinfo{pages}{19 } (\bibinfo{year}{2005}{\natexlab{a}}).

\bibitem[{\citenamefont{{De Raedt} et~al.}(2005{\natexlab{b}})\citenamefont{{De
  Raedt}, {De Raedt}, and Michielsen}}]{RAED05c}
\bibinfo{author}{\bibfnamefont{H.}~\bibnamefont{{De Raedt}}},
  \bibinfo{author}{\bibfnamefont{K.}~\bibnamefont{{De Raedt}}},
  \bibnamefont{and}
  \bibinfo{author}{\bibfnamefont{K.}~\bibnamefont{Michielsen}},
  \bibinfo{journal}{J. Phys. Soc. Jpn. Suppl.} \textbf{\bibinfo{volume}{76}},
  \bibinfo{pages}{16 } (\bibinfo{year}{2005}{\natexlab{b}}).

\bibitem[{\citenamefont{{De Raedt} et~al.}(2005{\natexlab{c}})\citenamefont{{De
  Raedt}, {De Raedt}, and Michielsen}}]{RAED05d}
\bibinfo{author}{\bibfnamefont{H.}~\bibnamefont{{De Raedt}}},
  \bibinfo{author}{\bibfnamefont{K.}~\bibnamefont{{De Raedt}}},
  \bibnamefont{and}
  \bibinfo{author}{\bibfnamefont{K.}~\bibnamefont{Michielsen}},
  \bibinfo{journal}{Europhys. Lett.} \textbf{\bibinfo{volume}{69}},
  \bibinfo{pages}{861 } (\bibinfo{year}{2005}{\natexlab{c}}).

\bibitem[{\citenamefont{Michielsen et~al.}(2005)\citenamefont{Michielsen, {De
  Raedt}, and {De Raedt}}}]{MICH05}
\bibinfo{author}{\bibfnamefont{K.}~\bibnamefont{Michielsen}},
  \bibinfo{author}{\bibfnamefont{K.}~\bibnamefont{{De Raedt}}},
  \bibnamefont{and} \bibinfo{author}{\bibfnamefont{H.}~\bibnamefont{{De
  Raedt}}}, \bibinfo{journal}{J. Comput. Theor. Nanosci.}
  \textbf{\bibinfo{volume}{2}}, \bibinfo{pages}{227 } (\bibinfo{year}{2005}).

\bibitem[{\citenamefont{{De Raedt} et~al.}(2006{\natexlab{a}})\citenamefont{{De
  Raedt}, {De Raedt}, and Michielsen}}]{RAED06z}
\bibinfo{author}{\bibfnamefont{H.}~\bibnamefont{{De Raedt}}},
  \bibinfo{author}{\bibfnamefont{K.}~\bibnamefont{{De Raedt}}},
  \bibnamefont{and}
  \bibinfo{author}{\bibfnamefont{K.}~\bibnamefont{Michielsen}}, in
  \emph{\bibinfo{booktitle}{Computer Simulation Studies in Condensed-Matter
  Physics XVIII}}, edited by \bibinfo{editor}{\bibfnamefont{D.~P.}
  \bibnamefont{Landau}}, \bibinfo{editor}{\bibfnamefont{S.~P.}
  \bibnamefont{Lewis}}, \bibnamefont{and} \bibinfo{editor}{\bibfnamefont{H.~B.}
  \bibnamefont{Sch{\"u}ttler}} (\bibinfo{publisher}{Springer},
  \bibinfo{address}{Berlin}, \bibinfo{year}{2006}{\natexlab{a}}), vol.
  \bibinfo{volume}{105}.

\bibitem[{\citenamefont{Michielsen et~al.}(2006)\citenamefont{Michielsen, {De
  Raedt}, and {De Raedt}}}]{MICH06z}
\bibinfo{author}{\bibfnamefont{K.}~\bibnamefont{Michielsen}},
  \bibinfo{author}{\bibfnamefont{H.}~\bibnamefont{{De Raedt}}},
  \bibnamefont{and} \bibinfo{author}{\bibfnamefont{K.}~\bibnamefont{{De
  Raedt}}}, in \emph{\bibinfo{booktitle}{Computer Simulation Studies in
  Condensed-Matter Physics XVIII}}, edited by
  \bibinfo{editor}{\bibfnamefont{D.~P.} \bibnamefont{Landau}},
  \bibinfo{editor}{\bibfnamefont{S.~P.} \bibnamefont{Lewis}}, \bibnamefont{and}
  \bibinfo{editor}{\bibfnamefont{H.~B.} \bibnamefont{Sch{\"u}ttler}}
  (\bibinfo{publisher}{Springer}, \bibinfo{address}{Berlin},
  \bibinfo{year}{2006}), vol. \bibinfo{volume}{105}.

\bibitem[{\citenamefont{{De Raedt} et~al.}(2006{\natexlab{b}})\citenamefont{{De
  Raedt}, Keimpema, {De Raedt}, Michielsen, and Miyashita}}]{RAED06c}
\bibinfo{author}{\bibfnamefont{K.}~\bibnamefont{{De Raedt}}},
  \bibinfo{author}{\bibfnamefont{K.}~\bibnamefont{Keimpema}},
  \bibinfo{author}{\bibfnamefont{H.}~\bibnamefont{{De Raedt}}},
  \bibinfo{author}{\bibfnamefont{K.}~\bibnamefont{Michielsen}},
  \bibnamefont{and}
  \bibinfo{author}{\bibfnamefont{S.}~\bibnamefont{Miyashita}},
  \bibinfo{journal}{Euro. Phys. J. B} \textbf{\bibinfo{volume}{53}},
  \bibinfo{pages}{139 } (\bibinfo{year}{2006}{\natexlab{b}}).

\bibitem[{\citenamefont{{De Raedt} et~al.}(2007{\natexlab{a}})\citenamefont{{De
  Raedt}, {De Raedt}, Michielsen, Keimpema, and Miyashita}}]{RAED07a}
\bibinfo{author}{\bibfnamefont{H.}~\bibnamefont{{De Raedt}}},
  \bibinfo{author}{\bibfnamefont{K.}~\bibnamefont{{De Raedt}}},
  \bibinfo{author}{\bibfnamefont{K.}~\bibnamefont{Michielsen}},
  \bibinfo{author}{\bibfnamefont{K.}~\bibnamefont{Keimpema}}, \bibnamefont{and}
  \bibinfo{author}{\bibfnamefont{S.}~\bibnamefont{Miyashita}},
  \bibinfo{journal}{J. Phys. Soc. Jpn.} \textbf{\bibinfo{volume}{76}},
  \bibinfo{pages}{104005} (\bibinfo{year}{2007}{\natexlab{a}}).

\bibitem[{\citenamefont{{De Raedt} et~al.}(2007{\natexlab{b}})\citenamefont{{De
  Raedt}, {De Raedt}, Michielsen, Keimpema, and Miyashita}}]{RAED07c}
\bibinfo{author}{\bibfnamefont{H.}~\bibnamefont{{De Raedt}}},
  \bibinfo{author}{\bibfnamefont{K.}~\bibnamefont{{De Raedt}}},
  \bibinfo{author}{\bibfnamefont{K.}~\bibnamefont{Michielsen}},
  \bibinfo{author}{\bibfnamefont{K.}~\bibnamefont{Keimpema}}, \bibnamefont{and}
  \bibinfo{author}{\bibfnamefont{S.}~\bibnamefont{Miyashita}},
  \bibinfo{journal}{J. Comp. Theor. Nanosci.} \textbf{\bibinfo{volume}{4}},
  \bibinfo{pages}{1} (\bibinfo{year}{2007}{\natexlab{b}}).

\bibitem[{MZI()}]{MZIdemo}
\bibinfo{note}{{Sample Fortran and Java programs and interactive programs that
  perform event-based simulations of a beam splitter, one Mach-Zehnder
  interferometer, and two chained Mach-Zehnder interferometers can be found at
  \url{http://www.compphys.net/dlm}}.}

\bibitem[{\citenamefont{Born and Wolf}(1964)}]{BORN64}
\bibinfo{author}{\bibfnamefont{M.}~\bibnamefont{Born}} \bibnamefont{and}
  \bibinfo{author}{\bibfnamefont{E.}~\bibnamefont{Wolf}},
  \emph{\bibinfo{title}{Principles of Optics}} (\bibinfo{publisher}{Pergamon},
  \bibinfo{address}{Oxford}, \bibinfo{year}{1964}).

\bibitem[{Qua()}]{QuantumTheory}
\bibinfo{note}{{We make a distinction between quantum theory and quantum
  physics. We use the term {\sl quantum theory} when we refer to the
  mathematical formalism, i.e., the postulates of quantum mechanics (with or
  without the wave function collapse postulate)~\cite{BALL03} and the rules
  (algorithms) to compute the wave function. The term {\sl quantum physics} is
  used for microscopic, experimentally observable phenomena that do not find an
  explanation within the mathematical framework of classical mechanics.}}

\bibitem[{\citenamefont{Baym}(1974)}]{BAYM74}
\bibinfo{author}{\bibfnamefont{G.}~\bibnamefont{Baym}},
  \emph{\bibinfo{title}{Lectures on Quantum Mechanics}}
  (\bibinfo{publisher}{W.A. Benjamin}, \bibinfo{address}{Reading MA},
  \bibinfo{year}{1974}).

\bibitem[{\citenamefont{Rarity and Tapster}(1997)}]{RARI97}
\bibinfo{author}{\bibfnamefont{J.~G.} \bibnamefont{Rarity}} \bibnamefont{and}
  \bibinfo{author}{\bibfnamefont{P.~R.} \bibnamefont{Tapster}},
  \bibinfo{journal}{Phil. Trans. R. Soc. Lond. A}
  \textbf{\bibinfo{volume}{355}}, \bibinfo{pages}{2267} (\bibinfo{year}{1997}).

\bibitem[{\citenamefont{Weihs et~al.}(1998)\citenamefont{Weihs, Jennewein,
  Simon, Weinfurther, and Zeilinger}}]{WEIH98}
\bibinfo{author}{\bibfnamefont{G.}~\bibnamefont{Weihs}},
  \bibinfo{author}{\bibfnamefont{T.}~\bibnamefont{Jennewein}},
  \bibinfo{author}{\bibfnamefont{C.}~\bibnamefont{Simon}},
  \bibinfo{author}{\bibfnamefont{H.}~\bibnamefont{Weinfurther}},
  \bibnamefont{and}
  \bibinfo{author}{\bibfnamefont{A.}~\bibnamefont{Zeilinger}},
  \bibinfo{journal}{Phys. Rev. Lett.} \textbf{\bibinfo{volume}{81}},
  \bibinfo{pages}{5039 } (\bibinfo{year}{1998}).

\bibitem[{\citenamefont{Adenier and Khrennikov}(2007)}]{ADEN07}
\bibinfo{author}{\bibfnamefont{G.}~\bibnamefont{Adenier}} \bibnamefont{and}
  \bibinfo{author}{\bibfnamefont{A.~Y.} \bibnamefont{Khrennikov}},
  \bibinfo{journal}{J. Phys. B: At. Mol. Opt. Phys.}
  \textbf{\bibinfo{volume}{40}}, \bibinfo{pages}{131 } (\bibinfo{year}{2007}).

\bibitem[{\citenamefont{Grimmet and Stirzaker}(1995)}]{GRIM95}
\bibinfo{author}{\bibfnamefont{G.~R.} \bibnamefont{Grimmet}} \bibnamefont{and}
  \bibinfo{author}{\bibfnamefont{D.~R.} \bibnamefont{Stirzaker}},
  \emph{\bibinfo{title}{Probability and Random Processes}}
  (\bibinfo{publisher}{Clarendon Press}, \bibinfo{address}{Oxford},
  \bibinfo{year}{1995}).

\bibitem[{\citenamefont{{De Raedt} et~al.}(2007{\natexlab{c}})\citenamefont{{De
  Raedt}, {De Raedt}, and Michielsen}}]{RAED07b}
\bibinfo{author}{\bibfnamefont{K.}~\bibnamefont{{De Raedt}}},
  \bibinfo{author}{\bibfnamefont{H.}~\bibnamefont{{De Raedt}}},
  \bibnamefont{and}
  \bibinfo{author}{\bibfnamefont{K.}~\bibnamefont{Michielsen}},
  \bibinfo{journal}{Comp. Phys. Comm.} \textbf{\bibinfo{volume}{176}},
  \bibinfo{pages}{642 } (\bibinfo{year}{2007}{\natexlab{c}}).

\bibitem[{WEI()}]{WEIHdownload}
\bibinfo{note}{\url{http://www.quantum.at/research/photonentangle/bellexp/data%
.html}. {The results presented here have been obtained by assuming that the
  data sets *\_V.DAT contain IEEE-8 byte (instead of 8-bit) double-precision
  numbers and that the least significant bit in *\_C.DAT specifies the position
  of the switch instead of the detector that fired.}}

\bibitem[{\citenamefont{Bell}(1993)}]{BELL93}
\bibinfo{author}{\bibfnamefont{J.~S.} \bibnamefont{Bell}},
  \emph{\bibinfo{title}{Speakable and unspeakable in quantum mechanics}}
  (\bibinfo{publisher}{Cambridge University Press},
  \bibinfo{address}{Cambridge}, \bibinfo{year}{1993}).

\bibitem[{\citenamefont{Weihs}(2000)}]{WEIH00}
\bibinfo{author}{\bibfnamefont{G.}~\bibnamefont{Weihs}}, Ph.D. thesis,
  \bibinfo{school}{University of Vienna} (\bibinfo{year}{2000}),
  \bibinfo{note}{{\url{http://www.quantum.univie.ac.at/publications/thesis/gwd%
iss.pdf}}}.

\bibitem[{\citenamefont{Kocher and Commins}(1967)}]{KOCH67}
\bibinfo{author}{\bibfnamefont{C.~A.} \bibnamefont{Kocher}} \bibnamefont{and}
  \bibinfo{author}{\bibfnamefont{E.~D.} \bibnamefont{Commins}},
  \bibinfo{journal}{Phys. Rev. Lett.} \textbf{\bibinfo{volume}{18}},
  \bibinfo{pages}{575 } (\bibinfo{year}{1967}).

\bibitem[{\citenamefont{Clauser and Horne}(1974)}]{CLAU74}
\bibinfo{author}{\bibfnamefont{J.~F.} \bibnamefont{Clauser}} \bibnamefont{and}
  \bibinfo{author}{\bibfnamefont{M.~A.} \bibnamefont{Horne}},
  \bibinfo{journal}{Phys. Rev. D} \textbf{\bibinfo{volume}{10}},
  \bibinfo{pages}{526 } (\bibinfo{year}{1974}).

\bibitem[{\citenamefont{Cirel'son}(1980)}]{CIRE80}
\bibinfo{author}{\bibfnamefont{B.~S.} \bibnamefont{Cirel'son}},
  \bibinfo{journal}{Lett. Math. Phys.} \textbf{\bibinfo{volume}{4}},
  \bibinfo{pages}{93 } (\bibinfo{year}{1980}).

\bibitem[{\citenamefont{Freedman and Clauser}(1972)}]{FREE72}
\bibinfo{author}{\bibfnamefont{S.~J.} \bibnamefont{Freedman}} \bibnamefont{and}
  \bibinfo{author}{\bibfnamefont{J.~F.} \bibnamefont{Clauser}},
  \bibinfo{journal}{Phys. Rev. Lett.} \textbf{\bibinfo{volume}{28}},
  \bibinfo{pages}{938 } (\bibinfo{year}{1972}).

\bibitem[{\citenamefont{Aspect et~al.}(1982{\natexlab{a}})\citenamefont{Aspect,
  Dalibard, and Roger}}]{ASPE82b}
\bibinfo{author}{\bibfnamefont{A.}~\bibnamefont{Aspect}},
  \bibinfo{author}{\bibfnamefont{J.}~\bibnamefont{Dalibard}}, \bibnamefont{and}
  \bibinfo{author}{\bibfnamefont{G.}~\bibnamefont{Roger}},
  \bibinfo{journal}{Phys. Rev. Lett.} \textbf{\bibinfo{volume}{49}},
  \bibinfo{pages}{1804 } (\bibinfo{year}{1982}{\natexlab{a}}).

\bibitem[{\citenamefont{Tapster et~al.}(1994)\citenamefont{Tapster, Rarity, and
  Owens}}]{TAPS94}
\bibinfo{author}{\bibfnamefont{P.~R.} \bibnamefont{Tapster}},
  \bibinfo{author}{\bibfnamefont{J.~G.} \bibnamefont{Rarity}},
  \bibnamefont{and} \bibinfo{author}{\bibfnamefont{P.~C.~M.}
  \bibnamefont{Owens}}, \bibinfo{journal}{Phys. Rev. Lett.}
  \textbf{\bibinfo{volume}{73}}, \bibinfo{pages}{1923 } (\bibinfo{year}{1994}).

\bibitem[{\citenamefont{Tittel et~al.}(1998)\citenamefont{Tittel, Brendel,
  Zbinden, and Gisin}}]{TITT98}
\bibinfo{author}{\bibfnamefont{W.}~\bibnamefont{Tittel}},
  \bibinfo{author}{\bibfnamefont{J.}~\bibnamefont{Brendel}},
  \bibinfo{author}{\bibfnamefont{H.}~\bibnamefont{Zbinden}}, \bibnamefont{and}
  \bibinfo{author}{\bibfnamefont{N.}~\bibnamefont{Gisin}},
  \bibinfo{journal}{Phys. Rev. Lett.} \textbf{\bibinfo{volume}{81}},
  \bibinfo{pages}{3563 } (\bibinfo{year}{1998}).

\bibitem[{\citenamefont{Rowe et~al.}(2001)\citenamefont{Rowe, Kielpinski,
  Meyer, Sackett, Itano, Monroe, and Wineland}}]{ROWE01}
\bibinfo{author}{\bibfnamefont{M.~A.} \bibnamefont{Rowe}},
  \bibinfo{author}{\bibfnamefont{D.}~\bibnamefont{Kielpinski}},
  \bibinfo{author}{\bibfnamefont{V.}~\bibnamefont{Meyer}},
  \bibinfo{author}{\bibfnamefont{C.~A.} \bibnamefont{Sackett}},
  \bibinfo{author}{\bibfnamefont{W.~M.} \bibnamefont{Itano}},
  \bibinfo{author}{\bibfnamefont{C.}~\bibnamefont{Monroe}}, \bibnamefont{and}
  \bibinfo{author}{\bibfnamefont{D.~J.} \bibnamefont{Wineland}},
  \bibinfo{journal}{Nature} \textbf{\bibinfo{volume}{401}}, \bibinfo{pages}{791
  } (\bibinfo{year}{2001}).

\bibitem[{\citenamefont{Fatal et~al.}(2004)\citenamefont{Fatal, Diamanti,
  Inoue, and Yamamoto}}]{FATA04}
\bibinfo{author}{\bibfnamefont{D.}~\bibnamefont{Fatal}},
  \bibinfo{author}{\bibfnamefont{E.}~\bibnamefont{Diamanti}},
  \bibinfo{author}{\bibfnamefont{K.}~\bibnamefont{Inoue}}, \bibnamefont{and}
  \bibinfo{author}{\bibfnamefont{Y.}~\bibnamefont{Yamamoto}},
  \bibinfo{journal}{Phys. Rev. Lett.} \textbf{\bibinfo{volume}{92}},
  \bibinfo{pages}{037904} (\bibinfo{year}{2004}).

\bibitem[{\citenamefont{{De Raedt} et~al.}(2007{\natexlab{d}})\citenamefont{{De
  Raedt}, Michielsen, Miyashita, and Keimpema}}]{RAED07d}
\bibinfo{author}{\bibfnamefont{H.}~\bibnamefont{{De Raedt}}},
  \bibinfo{author}{\bibfnamefont{K.}~\bibnamefont{Michielsen}},
  \bibinfo{author}{\bibfnamefont{S.}~\bibnamefont{Miyashita}},
  \bibnamefont{and} \bibinfo{author}{\bibfnamefont{K.}~\bibnamefont{Keimpema}},
  \bibinfo{journal}{Euro. Phys. J. B} \textbf{\bibinfo{volume}{58}},
  \bibinfo{pages}{55 } (\bibinfo{year}{2007}{\natexlab{d}}).

\bibitem[{\citenamefont{Aspect et~al.}(1982{\natexlab{b}})\citenamefont{Aspect,
  Grangier, and Roger}}]{ASPE82a}
\bibinfo{author}{\bibfnamefont{A.}~\bibnamefont{Aspect}},
  \bibinfo{author}{\bibfnamefont{P.}~\bibnamefont{Grangier}}, \bibnamefont{and}
  \bibinfo{author}{\bibfnamefont{G.}~\bibnamefont{Roger}},
  \bibinfo{journal}{Phys. Rev. Lett.} \textbf{\bibinfo{volume}{49}},
  \bibinfo{pages}{91 } (\bibinfo{year}{1982}{\natexlab{b}}).

\bibitem[{\citenamefont{Larsson and Gill}(2004)}]{LARS04}
\bibinfo{author}{\bibfnamefont{J.~A.} \bibnamefont{Larsson}} \bibnamefont{and}
  \bibinfo{author}{\bibfnamefont{R.~D.} \bibnamefont{Gill}},
  \bibinfo{journal}{Europhys. Lett.} \textbf{\bibinfo{volume}{67}},
  \bibinfo{pages}{707 } (\bibinfo{year}{2004}).

\bibitem[{\citenamefont{Santos}(2005)}]{SANT05}
\bibinfo{author}{\bibfnamefont{E.}~\bibnamefont{Santos}},
  \bibinfo{journal}{Phil. Mod. Phys.} \textbf{\bibinfo{volume}{36}},
  \bibinfo{pages}{544 } (\bibinfo{year}{2005}).

\bibitem[{\citenamefont{{\^Z}ukowksi}(2005)}]{ZUKO06}
\bibinfo{author}{\bibfnamefont{M.}~\bibnamefont{{\^Z}ukowksi}},
  \bibinfo{journal}{Stud. Hist. Phil. Mod. Phys.}
  \textbf{\bibinfo{volume}{36}}, \bibinfo{pages}{566} (\bibinfo{year}{2005}),
  \bibinfo{note}{{arXiv: quant-ph/0605034}}.

\bibitem[{\citenamefont{Sica}(1999)}]{SICA99}
\bibinfo{author}{\bibfnamefont{L.}~\bibnamefont{Sica}}, \bibinfo{journal}{Opt.
  Comm.} \textbf{\bibinfo{volume}{170}}, \bibinfo{pages}{55 }
  (\bibinfo{year}{1999}).

\bibitem[{\citenamefont{Hess and Philipp}(2004)}]{HESS04}
\bibinfo{author}{\bibfnamefont{K.}~\bibnamefont{Hess}} \bibnamefont{and}
  \bibinfo{author}{\bibfnamefont{W.}~\bibnamefont{Philipp}},
  \bibinfo{journal}{Proc. Natl. Acad. Sci. USA} \textbf{\bibinfo{volume}{101}},
  \bibinfo{pages}{1799 } (\bibinfo{year}{2004}).

\bibitem[{\citenamefont{Hess and Philipp}(2005)}]{HESS05b}
\bibinfo{author}{\bibfnamefont{K.}~\bibnamefont{Hess}} \bibnamefont{and}
  \bibinfo{author}{\bibfnamefont{W.}~\bibnamefont{Philipp}},
  \bibinfo{journal}{Found. of Phys.} \textbf{\bibinfo{volume}{35}},
  \bibinfo{pages}{1749 } (\bibinfo{year}{2005}).

\bibitem[{\citenamefont{Jaynes}(1989)}]{JAYN89}
\bibinfo{author}{\bibfnamefont{E.~T.} \bibnamefont{Jaynes}}, in
  \emph{\bibinfo{booktitle}{Maximum Entropy and Bayesian Methods}}, edited by
  \bibinfo{editor}{\bibfnamefont{J.}~\bibnamefont{Skilling}}
  (\bibinfo{publisher}{Kluwer Academic Publishers},
  \bibinfo{address}{Dordrecht}, \bibinfo{year}{1989}),
  vol.~\bibinfo{volume}{36}.

\bibitem[{\citenamefont{Tribus}(1999)}]{TRIB69}
\bibinfo{author}{\bibfnamefont{M.}~\bibnamefont{Tribus}},
  \emph{\bibinfo{title}{Rational Descriptions, Decisions and Designs}}
  (\bibinfo{publisher}{Expira Press}, \bibinfo{address}{Stockholm},
  \bibinfo{year}{1999}).

\bibitem[{\citenamefont{Jaynes}(2003)}]{JAYN03}
\bibinfo{author}{\bibfnamefont{E.~T.} \bibnamefont{Jaynes}},
  \emph{\bibinfo{title}{Probability Theory: The Logic of Science}}
  (\bibinfo{publisher}{Cambridge University Press},
  \bibinfo{address}{Cambridge}, \bibinfo{year}{2003}).

\bibitem[{\citenamefont{Cox}(1961)}]{COX61}
\bibinfo{author}{\bibfnamefont{R.~T.} \bibnamefont{Cox}},
  \emph{\bibinfo{title}{The Algebra of Probable Inference}}
  (\bibinfo{publisher}{Johns Hopkins University Press},
  \bibinfo{address}{Baltimore}, \bibinfo{year}{1961}).

\bibitem[{\citenamefont{{Zhao} and {De Raedt}}(2007)}]{ZHAO07b}
\bibinfo{author}{\bibfnamefont{S.}~\bibnamefont{{Zhao}}} \bibnamefont{and}
  \bibinfo{author}{\bibfnamefont{H.}~\bibnamefont{{De Raedt}}},
  \bibinfo{journal}{J. Comp. Theor. Nanosci.} \textbf{\bibinfo{volume}{(in
  press)}} (\bibinfo{year}{2007}).

\end{thebibliography}

\end{document}